\documentclass[prb,aps,amssymb,superscriptaddress,reprint]{revtex4-1}
\usepackage{graphicx}
\usepackage{dcolumn}
\usepackage{bm}
\usepackage{amsmath}
\usepackage{epstopdf}
\usepackage{natbib}
\usepackage{xcolor}
\usepackage{lineno,hyperref}

\newcommand{\Rmnum}[1]{\expandafter\@slowromancap\romannumeral #1@}
\makeatother
\begin{document}
\title{\bf Giant Topological Hall Effect in Magnetic Weyl Metal Mn$_{2}$Pd$_{0.5}$Ir$_{0.5}$Sn}

\author{ Arnab Bhattacharya$^\ddag$ }
\affiliation{Condensed Matter Physics Division, Saha Institute of Nuclear Physics, A CI of Homi Bhabha National Institute, 1/AF, Bidhannagar, Kolkata 700064, India}
\thanks{These authors contributed equally to this work.}
\author{ Sreeparvathy PC$^\ddag$ }
\affiliation{Department of Physics, Indian Institute of Technology Bombay, Mumbai 400076, India}
\thanks{These authors contributed equally to this work.}
\author{Afsar Ahmed}
\affiliation{Condensed Matter Physics Division, Saha Institute of Nuclear Physics, A CI of Homi Bhabha National Institute, 1/AF, Bidhannagar, Kolkata 700064, India}
\author{ Daichi Kurebayashi }
\affiliation{School of Physics, University of New South Wales, Sydney 2052, Australia}
\author{ Oleg A. Tretiakov }
\affiliation{School of Physics, University of New South Wales, Sydney 2052, Australia}
\author{Biswarup Satpati}
\affiliation{Condensed Matter Physics Division, Saha Institute of Nuclear Physics, A CI of Homi Bhabha National Institute, 1/AF, Bidhannagar, Kolkata 700064, India}
\author{Samik DuttaGupta}
\affiliation{Condensed Matter Physics Division, Saha Institute of Nuclear Physics, A CI of Homi Bhabha National Institute, 1/AF, Bidhannagar, Kolkata 700064, India}
\author{ Aftab Alam }
\email{aftab@iitb.ac.in}
\affiliation{Department of Physics, Indian Institute of Technology Bombay, Mumbai 400076, India}
\author{I. Das}
\email{indranil.das@saha.ac.in}
\affiliation{Condensed Matter Physics Division, Saha Institute of Nuclear Physics, A CI of Homi Bhabha National Institute, 1/AF, Bidhannagar, Kolkata 700064, India}
\def\thefootnote{$\ddag$}\footnotetext{These authors contributed equally to this work}
\begin{abstract}

The synergy between real and reciprocal space topology is anticipated to yield a diverse array of topological properties in quantum materials. We address this pursuit by achieving topologically safeguarded magnetic order in novel Weyl metallic Heusler alloy, Mn$_{2}$Pd$_{0.5}$Ir$_{0.5}$Sn. The system possesses non-centrosymmetric D$_{2d}$ crystal symmetry with notable spin-orbit coupling effects. Our first principles calculations confirm the topological non-trivial nature of band structure, including 42 pairs of Weyl nodes at/near the Fermi level, offering deeper insights into the observed anomalous Hall effect mediated by intrinsic Berry curvature. A unique canted magnetic ordering facilitates such rich topological features, manifesting through an exceptionally large topological Hall effect at low fields. The latter is sustained even at room temperature and compared with other known topological magnetic materials. Detailed micromagnetic simulations demonstrate the possible existence of an antiskyrmion lattice. Our results underscore the $D_{2d}$ Heusler magnets as a possible platform to explore the intricate interplay of non-trivial topology across real and reciprocal spaces to leverage a plethora of emergent properties for spintronic applications.

\end{abstract}

\maketitle

\section{Introduction}

\begin{figure*}[ht]
\begin{center}
\includegraphics[width=0.9\textwidth]{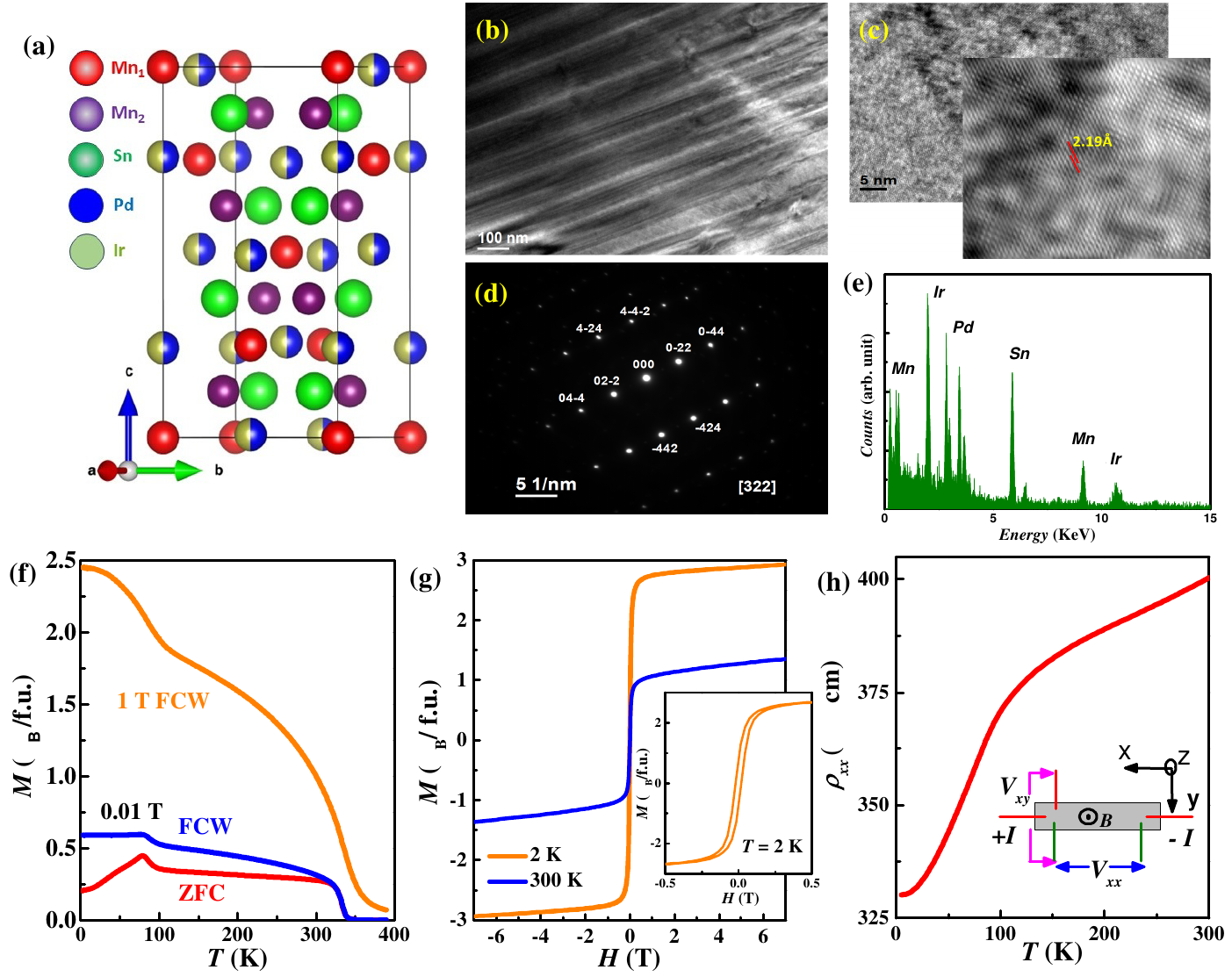}
\end{center}
\caption{{\bf(a)} Crystal structure of Mn$_{2}$Pd$_{0.5}$Ir$_{0.5}$Sn. {\bf(b)} TEM image of Mn$_{2}$Pd$_{0.5}$Ir$_{0.5}$Sn showing a twinned microstructure. {\bf(c)} High-resolution transmission electron microscopy image, inset shows Fourier filtered image {\bf(d)} Selected-area diffraction pattern showing the [322] orientation. {\bf(e)} EDX spectrum {\bf(f)} Thermomagnetic $M(T)$ curve under zero field cool and field cooled protocols. {\bf(g)} \textcolor{black}{Iso-thermal magnetization $M(H)$} at 2 and 300 K. Inset shows the low field hysteresis at 2 K. {\bf(h)} Temperature variation of $\rho_{xx}$ reflecting \textcolor{black}{the spin reorientation temperature, $T_{SR}$}. Schematics represent the orientation of the sample to the field i.e. applied perpendicular to length $\times$ breadth surface area (along z-axis). This arrangement is used for all magnetic and magnetotransport measurements.}
{\label{f1}}
\end{figure*}

Recent discovery of magnetic Weyl semimetals/metals (MWS) marked a significant leap in investigating an extraordinary electronic phase of matter, featuring non-trivial band intersections near the Fermi surface in reciprocal space ($\mathcal{KS}$)\cite{p85,p58}. This unveiled an array of intriguing magnetotransport properties within MWS, triggering intensified research\cite{p58,p19,p21}. MWS, characterized by broken time-reversal  ($\mathcal{T}$) and/or inversion symmetry($\mathcal{P}$), offer versatile means to adjust isolated Weyl nodes of opposite chirality in $\mathcal{KS}$ through magnetization ($M$)\cite{p49}. These Weyl points generate notable Berry curvature from occupied states, establishing a topological basis for the anomalous Hall effect (AHE) and hinting at potential advancements in topotronic devices\cite{p21,p22,p23,p48,p24,p50}. In contrast, real-space ($\mathcal{RS}$) topological magnetic spin configurations, such as skyrmions (Sk) and antiskyrmions (aSk), introduce an additional element in transverse resistivity $\rho_{xy}$, exhibiting a non-linear relationship with $M$. In metals, when conduction electrons pass through Sk/aSk, they accumulate non-zero Berry curvature due to an emergent magnetic field, $B_{em}= n_{sk}\cdot\phi_0$, where $n_{sk}$ and $\phi_0= h/e$ denotes the density and the flux quanta associated with the spin winding of each Sk/aSk, while $h$ and $e$ are Plank's constant and electronic charge, respectively. This extra phase acquired manifests as a measurable component of $\rho_{xy}$, $\rho_{xy}^T$, termed topological Hall effect (THE) \cite{p41,p4,p38,p39,p14,p17,p43,p92,p3,p6,p56,p57}. Though there exists extensive parallel research individually reporting (i) AHE originating from topological features in $\mathcal{KS}$ for collinear commensurate ferromagnets \cite{p10,p11} and (ii) Sk/aSk mediated THE in independent materials\cite{p65,p3}, discovering novel materials hosting both $\mathcal{KS}$ topological traits and topologically protected incommensurate noncoplaner magnetic spin ordering remains elusive and rare\cite{p86,p88,p87,p89}.

Interweaving magnetic spin texture with charge degrees of freedom, MWS enable superior electrical manipulation of spin texture compared to conventional ferromagnets, contingent to their realization in precisely engineered materials, \textcolor{black}{presenting} an intriguing challenge \cite{p76,p77}. The emergence of Weyl points in $\mathcal{KS}$ hinges upon the absence of a symmetry resulting from the product of $\mathcal{T}$ and parity, \textit{i.e.,} systems with broken $\mathcal{P}$ \cite{p61}, which also serves as a key requirement for stabilizing magnetic Sk/aSk lattices \cite{p4}. In this context, Heusler alloys of $D_{2d}$ structural symmetry with multiple magnetic sublattices emerge as a rare promising platform to unify $\mathcal{KS}$ topological properties with $\mathcal{RS}$ magnetic textures and explore their interrelation. The lack of $\mathcal{P}$, coupled with tailored spin-orbit coupling \textcolor{black}{(SOC)} amplifies the spin torque between magnetic sublattices, fostering competition between symmetric exchange and anisotropic Dzyaloshinskii-Moriya (DM) interaction. This leads to the stabilization of non-collinear spin textures. Moreover, the intrinsic magnetism acts as a Zeeman coupling, shifting the Weyl nodes in $\mathcal{KS}$, and consequently establishes a stable topological phase \cite{p9,p64,p75,Potkina2020,p68,p79}.

Employing the structural adaptability of Heusler alloys by chemical substitution, we introduce a pristine Weyl metal candidate, the tetragonal inverse Heusler alloy Mn$_{2}$Pd$_{0.5}$Ir$_{0.5}$Sn, engineered by doping half Palladium sites by Iridium in cubic Weyl metal Mn$_{2}$PdSn\cite{p105}. We demonstrate the concurrent presence of significant intrinsic AHE and extensive THE at low temperatures, persisting remarkably up to room temperature, and compare the same with other Sk/aSk based materials. A detailed \textit{ab-initio} and micromagnetic calculations affirm the presence of Weyl nodes at/near the Fermi level in conjunction with the aSk lattice phase, corroborating our experimental results. Our experimental findings substantiated by concomitant theoretical calculations hint at the co-existence of both real and reciprocal space topological phases in a single material, promising for the development of future topological magnetic devices.

\begin{figure*}[t]
\begin{center}
\includegraphics[width=.9\textwidth]{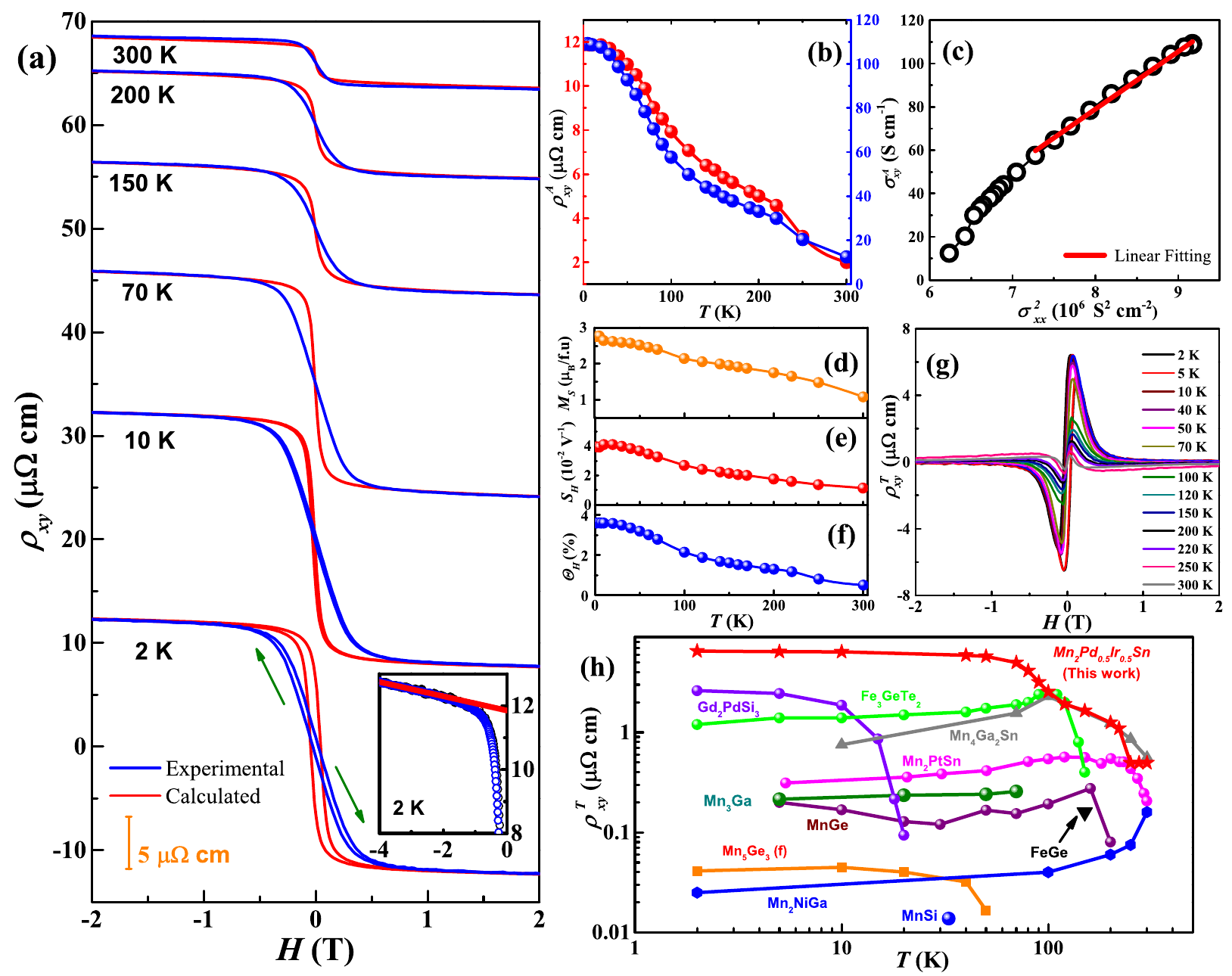}
\end{center}
\caption{\textbf{(a)} \textcolor{black}{Total Hall resistivity $\rho_{xy}$ (blue curve) and fitted sum of ordinary and anomalous Hall components (red curve) from 2 K to 300 K.} The arrows indicate the direction of hysteresis of $\rho_{xy}$. Inset shows the zero field extrapolation from the high field region of $\rho_{xy}$ to obtain the spontaneous zero field $\rho_{xy}^A$. \textbf{(b)} Temperature ($T$) variation of anomalous Hall resistivity and anomalous Hall conductivity $\sigma_{xy}^A$. \textbf{(c)} The anomalous Hall conductivity $\sigma_{xy}^A$ is plotted with $\sigma_{xx}^2$ and fits to $TYJ$ relation. \textbf{(d)}, \textbf{(e)} and \textbf{(f)} Temperature variation of saturation magnetization ($M_S$), anomalous Hall factor ($S_H$) and anomalous Hall angle ($\Theta_H$). \textbf{(g)} \textcolor{black}{Topological Hall resistivity} $\rho_{xy}^T (H)$ at different temperatures between 2 K to 300 K. \textbf{(h)} Comparing THE of Mn$_{2}$Pd$_{0.5}$Ir$_{0.5}$Sn with previous reports\cite{p42,p16,p3,p25,p41,p57,p80,p81}.}
{\label{f2}}
\end{figure*}

\section[ER]{Experimental Results}

Figure \ref{f1}(a) shows the crystal structure of Mn$_{2}$Pd$_{0.5}$Ir$_{0.5}$Sn, as derived from the Rietveld refinement of the room temperature X-ray diffraction  pattern of its polycrystalline sample with Cu-\textit{K}$_{\alpha}$ radiation (see Fig.S1 and Table ST1 of Supplementary Material (SM)\cite{R30}). The system crystallizes in a non-centrosymmetric inverse tetragonal Heusler structure with $D_{2d}$ crystal symmetry, ($I\bar{4}2d$, group no: 122). Figure \ref{f1}(b) shows the transmission electron microscopy (TEM) image showing high density of twinned microstructures. High resolution TEM (HRTEM) image in Fig. \ref{f1}(c) and inset therein show (204) lattice plane with lattice spacing of 2.19 $\AA$. The selected area electron diffraction (SAED) pattern from an arbitrary area is shown in Fig. \ref{f1}(d), which indicates the single crystalline nature of the material. SAED was indexed using the lattice parameters of the tetragonal crystal structure showing that the lamella was [322] oriented. To investigate the chemical composition of the compound, we have performed energy dispersive X-ray spectroscopy (EDX) in high-angle annular dark field (HAADF) scanning transmission electron microscopy (STEM-HAADF), verifying uniformity within $\sim 3\%$ of stoichiometric composition throughout the sample.

Figure \ref{f1}(f) illustrates the thermomagnetic $M(T)$ curves in zero-field cooled (ZFC) and field-cooled (FC) mode under various applied fields ($H$). Mn$_{2}$Pd$_{0.5}$Ir$_{0.5}$Sn undergoes a para to ferro (FM)/ferrimagnetic (FiM) transition at $T_C$ = 333 K. By further lowering the temperature, $M(T)$ shows a spontaneous increase at $T_{SR}$ = 85 K, consistent up to $H=$ 1 T. It can be argued that the presence of two interpenetrating non-equivalent Mn sublattices across distinct structural planes exposes them to differing environments. Earlier reports of isostructural Mn-based Heusler systems suggest that a competition arises between parallel and antiparallel exchange interactions, leading to a non-collinear canted ordering amid these two sublattices and evolving into a canted ferrimagnet along the easy axis\cite{p1,p6,p27,p70}. This results in the development of a net non-zero $ab$-plane magnetization which gradually gets suppressed with the increase in $T$ until a collinear ferrimagnetic ordering is established above $T_{SR}$. Figure \ref{f1}(g) shows the isothermal magnetization $M(H)$ at $T=$ 2 and 300 K. The system attains a saturation moment $M_S$ of 2.37 $\mu_B/f.u.$ at 2 K accompanied by a small hysteresis (see inset Fig. \ref{f1}(g)). However, this hysteresis diminishes gradually as the temperature increases. \textcolor{black}{This significantly enhanced magnetization, relative to other Sk/aSk-hosting systems based on transition elements \cite{p17,p56,p71,p3}, coupled with notable SOC effects due to high-$Z$ elements and the non-centrosymmetric crystal symmetry, aids in stabilizing topological spin textures.} The longitudinal resistivity $\rho_{xx}(T)$ exhibits metallic behaviour, aligning consistently with the spin-reorientation transition observed in thermomagnetic measurements (Fig\ref{f1}(h)). This establishes an intricate correlation between magnetic ordering and transport properties, prompting us to examine the magnetotransport properties of this system, aiming to uncover possible topological magnetic spin textures.

\textcolor{black}{Figure \ref{f2}(a) depicts the field variation of $\rho_{xy}$, mirroring the $M(H)$ isotherms. This, coupled with effective magnetocrystalline anisotropy (Fig. S2 in SM\cite{R30}) and minimal magnetoresistance (Fig. S3 in SM\cite{R30}), is suggestive of the predominant contribution of AHE in $\rho_{xy}$.} It is worth noting, despite qualitatively resembling the magnetic isotherms, $\rho_{xy}(H)$ exhibits an opposite hysteresis at lower temperatures. For FM/FiM systems with non-collinear spin arrangements, $\rho_{xy}$ is empirically expressed as, $\rho_{xy} = \rho_{xy}^N + \rho_{xy}^A + \rho_{xy}^T$, where $\rho^{N}_{xy} = R_0H$, $\rho^{A}_{xy} = R_SM$ corresponds to normal and anomalous  Hall resistivity, respectively \cite{p3,p6,p43,p92}. The negative slope of $\rho_{xy}$ affirms the electron as majority carriers with density $n_0 = 5.21 \times 10^{21}$ cm$^{-3}$ at $T = 2$ K, deduced by employing the relation $n_0 = -1/|e|R_0$. The normal contribution is estimated to be 1-3$\%$ of $\rho_{xy}$ in the 2-7 T field range where magnetization attains saturation. The $\rho_{xy}^A$ is estimated from the high field extrapolation of $\rho_{xy}(H)$ to zero field. Figure \ref{f2}(b) shows the temperature variation of $\rho_{xy}^A$ where it reaches the maximum saturated value of $\sim$ 11.88 $\mu\Omega$cm at $T=$ 2 K, presenting a unique observation among Mn-based Heusler alloys\cite{p3,p4,p6}. This prompted the investigation to unravel the underlying mechanism. To analyze this, we derive the anomalous Hall conductivity $\sigma_{xy}^A$ (AHC), akin to obtaining $\rho_{xy}^A$, from total Hall conductivity $\sigma_{xy} = \rho_{xy}/(\rho_{xy}^2 + \rho_{xx}^2)$. In a broader context, the AHE manifests through either intrinsic mechanisms, involving Berry curvature generated within $\mathcal{KS}$, or extrinsic mechanisms like side-jump ($sj$) and skew ($sk$) scattering\cite{p15}. The total AHC, $\sigma_{xy}^A$, can be expressed as a sum of three terms $\sigma_{xy}^A = \sigma_{sk}^A + \sigma_{sj}^A + \sigma_{int}^A$, where $\sigma_{sk}^A$, $\sigma_{sj}^A$ and $\sigma_{int}^A$ denote the skew, side-jump and intrinsic conductivity, respectively. Subsequently, we adopt the TYJ scaling relation for $\sigma_{xy}^A$ \textit{i.e.,} $\sigma_{xy}^A = -\kappa\sigma_{xx0}^{-1}\sigma_{xx}^2 - b =-a\sigma_{xx}^2 - b$, where $b = \rho_{xy}^A/\rho_{xx}^2$ correlates with $\sigma_{int}^{A}$ and $\sigma_{xx0}$ denotes residual longitudinal conductivity, to quantify the intrinsic and extrinsic contributions\cite{p2,p15,p20,p48,p93,p94,p95}. Figure \ref{f2}(c) shows the $\sigma_{xy}^A$ versus $\sigma_{xx}^2$ plot over the entire temperature range. From the linear fitting, we obtain an intrinsic Berry curvature contribution to AHC of $b=$ 133 S cm$^{-1}$. It is possible that the small deviation from the linearity in the high temperature region can be attributed to the broadening of Fermi-Dirac distribution with the increase of temperature (see Fig. S5 and also Fig. S6 for angle dependence of AHC)\cite{p96,p97,R30}. The obtained $\sigma_{int}^A$ turns out to be $\sim 0.42e^2/(hc)$ where $c$ is the lattice parameter\cite{98}. At lower temperatures with reduced phonon scattering, $\sigma_{sj}^A$ entangles with $\sigma_{xy, int}^A$, challenging differentiation due to lack of theoretical modelling. Though we can't precisely estimate $\sigma_{sj}^A$, we get an idea of its order of magnitude, employing the relation, $(e^2/(hc)(\epsilon_{SOC}/E_F)$, where $\epsilon_{SOC}$ and $E_F$ is the spin-orbit coupling strength and Fermi energy\cite{p74,p33}. For itinerant metals, $\epsilon_{SOC}/E_F \approx 0.01$ yields $\sigma_{xy,sj}^{A} \sim$ 3.14 S cm$^{-1}$, notably smaller than $\sigma_{int}^{A}$ confirming the predominant intrinsic mechanism in $\sigma_{xy}^{A}$. Figure \ref{f2}(d)-(f) illustrates the $T$ variation of saturation magnetization ($M_S$), anomalous Hall factor $S_H$(=$\sigma_{xy}^A/M$) and anomalous Hall angle (AHA, $\Theta_H = \sigma_{xy}^A/\sigma_{xx} (\%)$), respectively. It is clear that the temperature variation of $\sigma_{xy}^A$ and $M_S$ shares qualitative similarity. This behaviour is quantitatively reflected in the scale factor $S_H \approx 0.04$ V$^{-1}$ which remains almost constant, verifying the scattering-independent Karplus-L$\ddot{u}$ttinger origin of AHC\cite{p34}. The $\Theta_H$ attains a robust value of 3$\%$ at $T$ = 2 K, but moderately decreases at higher temperatures due to inelastic scattering\cite{p33,p74}. Therefore, both $\Theta_H$ and $S_H$ quantify the robust intrinsic origin of AHE. To delve deeper into the origin of the substantial intrinsic AHC, we investigated the band structure of Mn$_{2}$Pd$_{0.5}$Ir$_{0.5}$Sn by detailed \textit{ab-initio} calculation in Section \ref{FPC}.

\begin{figure*}[ht]
\centering
\includegraphics[width=0.95\linewidth]{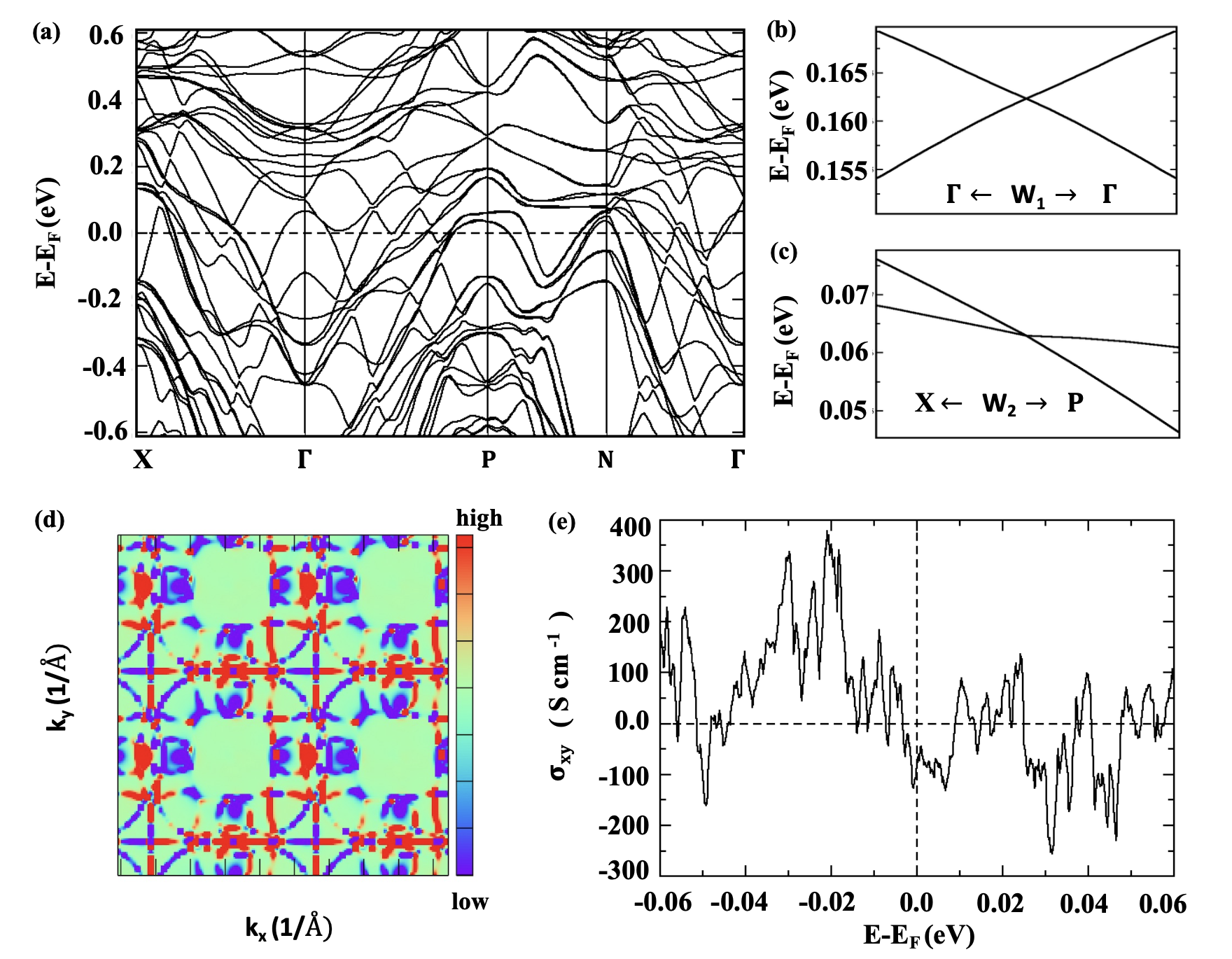}
\caption{(Color online) For Mn$_{2}$Pd$_{0.5}$Ir$_{0.5}$Sn {\bf (a)} Band structure with spin-orbit coupling. {\bf (b)} Band dispersion around one of the Weyl points (type I) {\bf (c)} Band dispersion around one of the Weyl points (type II) {\bf (d)} Berry curvature in $k_{x}$-$k_{y}$ plane with $k_{z} =0$  {\bf (e)} Energy dependence of anomalous Hall conductivity. }
\label{tfig2}
\end{figure*}

We speculate that the non-centrosymmetric crystal structure, magnetocrystalline anisotropy and the metallic nature of this system entail the possibility of realizing the additional contribution, $\rho_{xy}^T$ associated with the topological spin textures. Leveraging the intrinsic nature of AHE, we formulated the anomalous Hall coefficient $R_S = \gamma\rho_{xx}^2$, with $R_S$ exhibiting $H$ invariance stemming from observed minuscule magnetoresistance\cite{p14,p43,R30}. For $H\ge\pm$2 T, Mn$_{2}$Pd$_{0.5}$Ir$_{0.5}$Sn attains ferromagnetic saturated state where non-collinear spin states mediated scaler spin chirality $\chi_{ij}= \sum \hat{S_i}\cdot(\hat{S_j} \times \hat{S_k})$, $\hat{S_i}$ is the spin vector of magnetic moment, and $\rho_{xy}^T$ are absent. Hence we model the transverse resistivity as $(\rho_{xy}/H) = R_0 + \gamma(\rho_{xx}^2M/H)$. The $(\rho_{xy}/H)$ versus $(\rho_{xx}^2M/H)$ curve exhibits a good linear fit with the slope and intercept yield $R_0$ and $\gamma$, respectively (see Fig.S8 of SM for more detail\cite{R30}). \textcolor{black}{The calculated ($R_0H + \gamma \rho_{xx}^2M$) curve shows distinct deviation from $\rho_{xy}(H)$ in Fig.\ref{f2}(a), a feature attributed to topological magnetic ordering \cite{p121,p122,p119,p123}. We estimated $\rho_{xy}^T$, by subtracting the calculated curve from $\rho_{xy}(H)$ (in Fig.\ref{f2}(g)), which remarkably exhibit three intriguing features. First, an exceptionally large topological Hall effect $\rho_{xy}^T \sim 6.44 \mu\Omega$cm is observed at $T=$ 2 K ($\sigma_{xy}^T \approx \rho_{xy}^T/\rho_{xx}^2$ = 59.73 S cm$^{-1}$) with a hysteresis and the switching of $\rho_{xy}^T$ opposite to magnetization. This behaviour has recently been associated with magnetic aSks stabilized by magnetocrystalline anisotropy with the core antiparallel to applied field direction\cite{p3,p6,p43,p92}. }Second, the presence of significant $\rho_{xy}^T$ at $H =$0 T, demonstrating the formation of robust topological magnetic spin textures at zero field below $T = 15$ K. And third, the presence of significant $\rho_{xy}^T$ on either side of $T_{SR}$ even up to room temperature (Fig.S9 of SM\cite{R30}), a unique observation within Mn-based D$_{2d}$ Heusler family\cite{p3,p6,p56,p57}. \textcolor{black}{Accounting in the theory of THE\cite{p82,p124,p125,p126,p120}, to verify the validity of $\mathcal{RS}$ topological picture for THE on either side of $T_{SR}$, we scaled the maximum of topological Hall conductivity ($\sigma_{max}^T$) with $\sigma_{xx}$. The near quadratic variation of $\sigma_{max}^T$ with $\sigma_{xx}$ ($\propto \tau^2$) for increasing $\sigma_{xx}$, over the entire measured temperature range (Fig.S10 of SM\cite{R30}), attests to the $\mathcal{RS}$ topological ordering driven THE.}

For a more comprehensive analysis, we compare $|\rho_{max}^T|$ in Fig. \ref{f2}(e), with other Sk/aSk hosting materials such as the B20 compound \cite{p41,p80}, iso-symmetric Heusler alloys \cite{p3,p4,p6,p16,p81}, and geometrically frustrated system like Gd$_2$PdSi$_3$ \cite{p42}. Mn$_{2}$Pd$_{0.5}$Ir$_{0.5}$Sn notably outperforms all of these, even surpassing Gd$_2$PdSi$_3$ with a \textit{'Giant Topological Hall effect'}. The large observed THE with significant remnant $\rho_{xy}^T$ at $H=$ 0 T hints at the possible formation of ground-state topological spin textures, intriguing for memory device applications. Non-centrosymmetric $D_{2d}$ crystal symmetry ensures the stabilization of asymmetric DM interaction in basal ab-plane, $D_{x}=-D_{y}$, enhancing the possibility of aSk stabilization. It is noteworthy to mention, the magnetocrystalline anisotropy energy plays a crucial role in the realization of topological aSk spin configuration for tetragonal materials. The presence of considerable magnetic anisotropy can be inferred from the hysteresis loop present in out-of-plane $M(H)$ isotherm in Fig. \ref{f1}(g) at $T=2$ K \cite{p3}. To gain insights about creating and stabilising the possible aSk phase mediated by crystal geometry and the anisotropy energy, we performed a detailed micromagnetic simulation, as shown in Section \ref{MMC}.


\begin{figure*}[t]
\centering
\begin{center}
\includegraphics[width=0.9\textwidth]{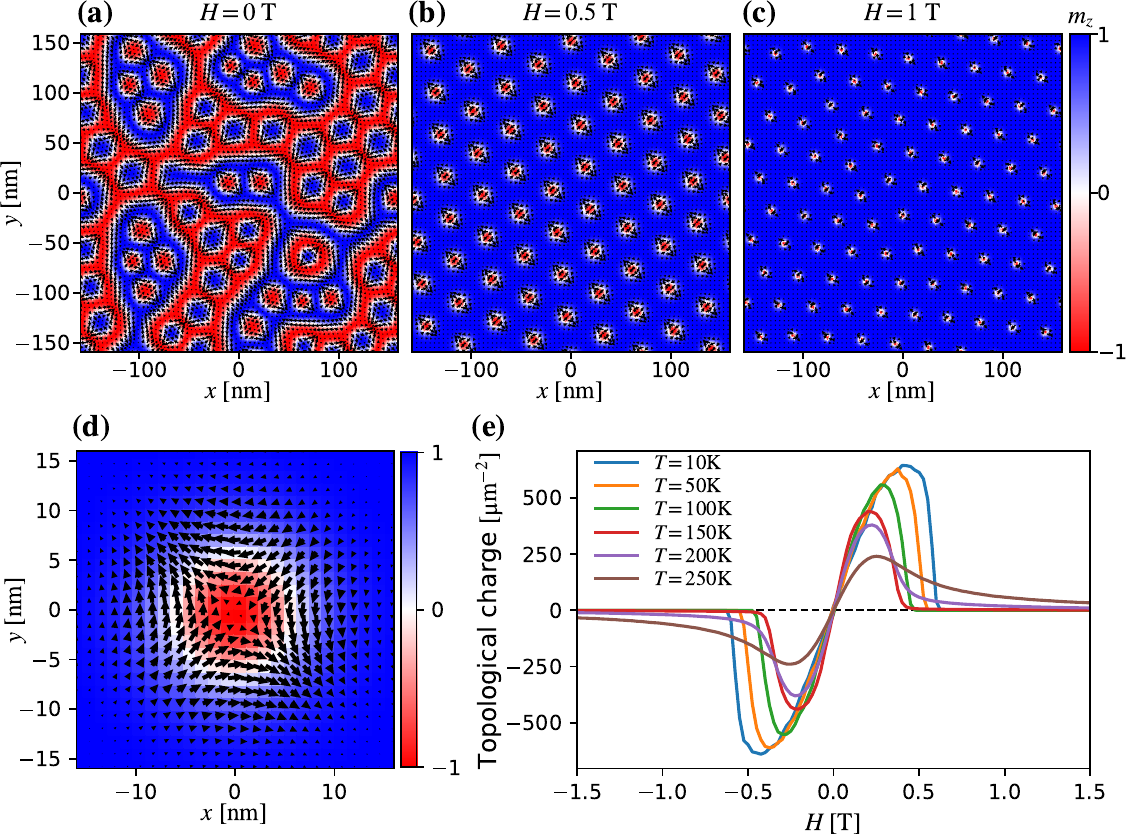}
\end{center}
\centering
\caption{\textbf{(a)-(c)} Magnetic configurations are obtained by micromagnetic simulations for various magnetic fields, $H = 0, 0.5, 1$T.
	The black arrows indicate the in-plane magnetic moments, while the color density shows the out-of-plane magnetic component.
	\textbf{(d)} The magnified image of an antiskyrmion depicts a detailed internal spin structure.
	\textbf{(e)} The topological charge density as a function of an out-of-plane magnetic field is shown for different temperatures. For the evaluation of the topological charge, $K=0.4\times 10^5\ \rm J/m^3$ is used.}
{\label{Topology}}
\end{figure*}

\section{First principles calculations} \label{FPC}

To gain further insights into the possible magnetic ground state of Mn$_{2}$Pd$_{0.5}$Ir$_{0.5}$Sn, and clarify the underlying physics responsible for the observed magnetotransport behaviour, we performed first-principles calculations using Vienna Ab-initio simulation package (VASP)\cite{p30,p31}. Further details about the computational methods are given at the end of this manuscript. Table ST4 of SM\cite{R30} summarizes the relative energies of various (both collinear and non-collinear) magnetic configurations along with their magnetic structures (see Figs. S12(a)-(k) of SM)\cite{R30}. The choice of these magnetic ordering is not random but has a certain background. Magnetic ordering of several Heusler alloys belonging to the same family as Mn$_{2}$Pd$_{0.5}$Ir$_{0.5}$Sn are investigated in detail using Neutron diffraction experiments and first-principles calculations \cite{p27,p6,p28,p1}. A few pertinent examples include, (1) A canted non-collinear magnetic configuration of Mn$_2$RhSn  belonging to the same tetragonal family with D$_{2d}$ point group. In this case, the magnetic moments of two in-equivalent Mn atoms, Mn$_{I}$ and Mn$_{II}$, are directed along the z-axis and 55$^0$ canted from the z-axis, respectively\cite{p27} (2) Mn$_2$PtIn and Mn$_2$IrSn show a similar pattern of magnetic configuration (see \cite{p28,p27}). (3) Mn$_x$Pt$_{0.9}$Pd$_{0.1}$Sn is yet another prototype candidate with space group $I\bar{4}2d$, reported to host a non-collinear magnetic phase from neutron diffraction measurement. In this system, magnetic moments of the two in-equivalent Mn atoms (Mn$_{I}$ and Mn$_{II}$) are aligned in an ab-plane, resulting in a ferrimagnetic ordering\cite{p6}. Because the magnetic atom (Mn) is responsible for dictating the magnetic ordering in all these systems, we took a hint from the above studies and simulated all those magnetic orderings in our system Mn$_{2}$Pd$_{0.5}$Ir$_{0.5}$Sn. All the configurations are fully relaxed. The relative energies ($\Delta$E) and magnetic moment directions for these numerous magnetic configurations are shown in Table ST4 of SM\cite{R30}. To make a comparison, Canted-2 is a similar magnetic configuration as those predicted for Mn$_2$RhSn system, Canted-3 is a similar magnetic ordering as predicted in Mn$_x$Pt$_{0.9}$Pd$_{0.1}$Sn. We believe that such an approach to scan a wide collection of magnetic configurations, which are already realized experimentally in similar prototype compounds (same family as Mn$_x$Pt$_{0.9}$Pd$_{0.1}$Sn) is a reasonable approach to evaluate the ground state magnetic ordering in our present system. We found that the Canted-1 state is the lowest-energy state, with lattice constants $a$ = $6.6$  \r{A} and $c$ = $11.62$ \r{A}. The experimental lattice constants are  $a$ = $6.35$ \r{A}, $c$ = 12.24 \r{A}, which agrees fairly well with the theoretically optimized ones.The computed non-collinear magnetic moments at  Mn$_{I}$ and Mn$_{II}$ atoms in the canted-1 configuration are (2.6, 0, -2.2) and (-2.4, 0, 2.2) $\mu$$_{B}$ respectively, with an angle  177.1$^{\circ}$ between them. Interestingly, few of the magnetic configurations are energetically very close to Canted-1, for instance, FiM-1 \& Canted-1; Canted-3 \& Canted-4. Such a small energy difference can possibly be attributed to the effect of DM interaction. 

Table \ref{Exchange} shows the strength of the nearest and higher neighbouring exchange interactions (J$_{ij}$) between different Mn-Mn pairs and the corresponding bond lengths between them. The NP exchange interaction, J$_{12}$, is the strongest among all the interactions with an antiferromagnetic nature, similar to that reported for other Heusler alloys e.g. Mn$_{2}$RhSn\cite{p27}. Despite the same bond length, the strength of the NNP interactions (Mn$_{I}$-Mn$_{I}$ and Mn$_{II}$-Mn$_{II}$) are different, which can be attributed to the proximity effect of Pd, Ir, and Sn atoms located closer to Mn$_{I}$ compared to Mn$_{II}$ atoms. As expected, the interaction strengths keep on decreasing for higher neighbours, becoming negligibly small for the fourth neighbouring pairs. The higher neighbour (beyond 2nd neighbour) interactions are ferromagnetic in nature. The observation of the competing nearest and next-nearest neighbor exchange interaction strength indicates the possibility of the stabilization of non-collinear magnetic structures in Mn$_{2}$Pd$_{0.5}$Ir$_{0.5}$Sn. In addition, we have also calculated the  DM interaction between the nearest neighbour Mn$_{I}$-Mn$_{II}$ pair. For this purpose, we have performed four constrained magnetic calculations, as illustrated in Ref. \cite{p32}.  D$_{12}$$^{x}$, D$_{12}$$^{y}$, and D$_{12}$$^{z}$ indicate the components of DM interaction vector D$_{12}$ along $x, y$ and $z$ directions with their corresponding values  -0.867 meV,  -0.009 meV, and 0.616 meV respectively, clearly showing its anisotropic nature. As shown later, this anisotropic DMI plays a key role in stabilising spin textures, leading to a large topological Hall effect. DM interactions die down rapidly to almost negligibly small values beyond first neighbouring pair.

\begin{table}[t]
\caption{Exchange interactions and bond lengths between different Mn-Mn pairs in Mn$_{2}$Pd$_{0.5}$Ir$_{0.5}$Sn}
\begin{ruledtabular}
\begin{tabular}{c c c}
Pairs & Exchange  Interaction(meV) & Bond length(\r{A})\\
\hline
\noalign{\smallskip}
\noalign{\smallskip}
\indent
J$_{12}$ \indent(NP) & -24.4 & 2.71 (Mn$_{I}$-Mn$_{II}$)  \\
J$_{11}$ (NNP) & -18.02  & 4.41 (Mn$_{I}$-Mn$_{II}$) \\
J$_{22}$ (NNP) &-13.2 & 4.41 (Mn$_{II}$-Mn$_{II}$)\\
J$_{11}^{\prime}$ (NNNP) & +10.06  & 4.49 (Mn$_{I}$-Mn$_{I}$) \\
J$_{22}^{\prime}$ (NNNP) &+2.16 & 4.49 (Mn$_{II}$-Mn$_{II}$)\\
J$_{12}^{\prime\prime}$ (NNNNP) &+0.002 & 5.11 (Mn$_{I}$-Mn$_{II}$)\\
\end{tabular}
\end{ruledtabular}
\label{Exchange}
\end{table}

Figure \ref{tfig2}(a) shows the band structure of Mn$_{2}$Pd$_{0.5}$Ir$_{0.5}$Sn in the lowest energy canted-1 magnetic configuration. One can notice several band crossings at/near the $E_{F}$, which can have topological Weyl characteristics because the system breaks both time reversal as well as inversion symmetry. A detailed examination of nodal points around E$_{F}$ revealed 42 pairs of Weyl points with $\pm$1 chirality. The {\bf k}-coordinates of these Weyl points are provided in Table ST5 of SM\cite{R30}. Band dispersion around one of the type-1 Weyl points (-0.249, -0.49, 0.0)  and type II Weyl points (0.26,0.00, 0.01) are shown in Fig. \ref{tfig2}(b) and \ref{tfig2}(c) respectively. These Weyl crossings yield finite Berry curvature calculated in the k$_x$-k$_y$ plane, as presented in Fig. \ref{tfig2}(d). Topological non-trivial features get reflected in the AHC via the contribution from intrinsic Berry curvature. Figure \ref{tfig2}(e) shows the energy-dependent AHC, the magnitude of which at/near E$_F$ is 78 S cm$^{-1}$, which is in fair agreement with our measured value. A similar comparison is also observed in several other Heusler alloys belonging to the same family as Mn$_{2}$Pd$_{0.5}$Ir$_{0.5}$Sn.  \cite{p105,p113,p114,p115,p116,p117,p118,p58,p11,p10} 

\textcolor{black}{For completeness, we have also checked the possible high entropy (HE) nature of Mn$_{2}$Pd$_{0.5}$Ir$_{0.5}$Sn, being a multi-component alloy. To do this, we have used the Partial occupation (POCC) algorithm proposed recently by Curtarolo et al.\cite{p128}. Within this algorithm, the structures are statistically weighted through appropriate Boltzmann factors and symmetry degeneracies to produce ensembles determining physical properties. We have simulated several descriptors, such as configurational, vibrational, electronic and magnetic entropy, and Vickers hardness, to evaluate the HE nature of   Mn$_{2}$Pd$_{0.5}$Ir$_{0.5}$Sn (see Fig. S13, Table ST6, ). Our detailed analysis suggests that the present system does not fall into the high entropy alloy category. For more details about the simulated results, see DFT Calculations section of SM\cite{R30}.}

\section{Micromagnetic simulations} \label{MMC}

Finally, to clarify the origin behind the observation of significant THE, we carry out micromagnetic simulations using the micromagnetic parameters determined from experiments and first-principle calculations. In our simulation, we used the experimental values of the saturation magnetization $M_s$ and the anisotropy constant $K$, while the exchange stiffness $A$ and the DMI constant $D$ are taken from the results of the first-principle calculations.

\begin{figure*}[t]
\centering
\includegraphics[width=\linewidth]{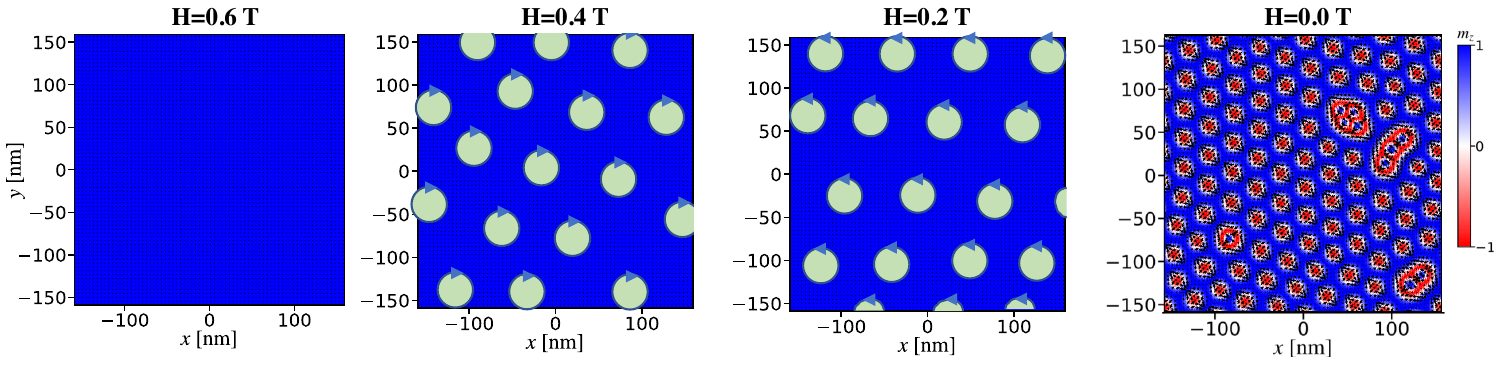}
\caption{\textcolor{black}{  Topological phase transition of Mn$_2$Pd$_{0.5}$Ir$_{0.5}$Sn in $\mathcal{R}$-space, displaying the evolution of spin textures with increasing the field ($H$) strength. The color map corresponds to the spatial distribution of the out-of-plane spin component ($m_z$).}}
\label{f5}
\end{figure*}

Figure \ref{Topology}(a)-(c) shows the micromagnetic snapshots of the magnetic configuration under applied field \textit{H} = 0, 0.5 and 1 T, respectively for fixed \textit{K} = 1.5$\times 10^5$ J$m^{-3}$ and \textit{D} = 0.693 mJ$m^{-2}$. We observe that under zero \textit{H}, the magnetic ground state is comprised of bubble-like domain patterns embedded within a uniaxial stripe background. This result is in good agreement with the experimentally observed remanent topological Hall resistivity, demonstrating that topological spin textures can be the ground state of Mn$_{2}$Pd$_{0.5}$Ir$_{0.5}$Sn\cite{p101}. Interestingly, the application of \textit{H}, results in the stabilization of an antiksyrmionic lattice, where the antiskyrmions decrease in size with the increase applied $H$. Figure \ref{Topology}(d) shows the enlarged view of the antiskyrmionic structure. Considering $\rho_{xy}^T = PR_0(h/ea_{sk}^2)$, where $P$ and $a_{sk}$ are electron polarization for conduction electron and size of the topological spin texture, respectively, we obtain $a_{sk} \sim$ 10 nm at \textit{T} = 2 K, corroborating with the simulation results. 
Figure \ref{Topology}(e) shows the topological charge density \textcolor{black}{$q(\boldsymbol{r})$} under the application of \textit{H} for various \textit{T}. 
\textcolor{black}{
Within the semiclassical transport theory \cite{p124,p125,p126,p120}, the topological Hall conductivity is directly proportional to the topological charge density~\cite{p127}; 
$
\sigma^T_{xy} = h \tau / (2 m) \sigma_{xx} \langle q(\mathbf{r})\rangle,
$
where $h$ is the Plank constant, $\tau$ is a scattering lifetime, $m$ is an electron mass, $\sigma_{xx}$ is a longitudinal conductivity, and $\langle q(\mathbf{r})\rangle$ is a spatial average of the topological charge density.
}
Clearly, the topological charge density attains a maximum with decreasing \textit{T}, in qualitative agreement with the experimental results\textcolor{black}{, as presented in Fig.~\ref{f2} (g)}. 
This indicates that the topological Hall resistivity primarily originates from the formation of an antiskyrmion lattice. To demonstrate the robustness of our micromagnetic results for a wider range of material parameters (and longer ranged interactions, if any), we have calculated the micromagnetic landscapes for few other DM interactions (\textit{D} = 0.347, 0.69 and 1.04 mJ$m^{-2}$) and anisotropy values (\textit{K} = 1.2$\times 10^5$ J$m^{-3}$ and 2.0$\times 10^5$ J$m^{-3}$), as shown in Figs. S14 and S15 of SM\cite{R30} respectively. Clearly the antiskyrmion lattices remain robust for other material's parameters.

\textcolor{black}{Several of these aSk lattices show periodic patterns (e.g. Fig. \ref{Topology}(b,c) and lower panels of Fig. S14 and S15). If one considers these aSk as quasi-particles, these periodically arranged quasi-particles can have non-trivial band-dispersions similar to Fermionic electrons, resulting in a nontrivial Berry curvature stemming from the intricate interplay between $\mathcal{RS}$ and $\mathcal{KS}$ Berry curvature. Such interconnection can be captured within the tight-binding (TB) model of the aSk lattice, presented in the following section\cite{p67}.}

\section{Tight-binding model for antiskyrmion}

\textcolor{black}{The TB-method used here to understand the interplay between real and k-space topology of Mn$_2$Pd$_{0.5}$Ir$_{0.5}$Sn is similar to that employed in Refs. \cite{p67,p139}. As such, to avoid redundancy, we refer the readers to these references for the methodology details and provide the salient features and the corresponding key terminologies here.}

\textcolor{black}{It is quite interesting to find that the multiple-space topology cross-over can be achieved in Mn$_2$Pd$_{0.5}$Ir$_{0.5}$Sn  by tuning external fields. For this we choose four different external fields, $H$=0, 0.2, 0.4 and 0.6 T, to calculate their $\mathcal{K}$-space edge states and $\mathcal{R}$-space spin textures. Figure \ref{f5} shows the topological phase transition of Mn$_2$Pd$_{0.5}$Ir$_{0.5}$Sn in $\mathcal{R}$-space. We found that when $H$=0.6 T (and above), Mn$_2$Pd$_{0.5}$Ir$_{0.5}$Sn  exhibits the trivial FM configuration, as also observed from our micromagnetic simulation. For any field less than 0.5 T, different nature of topological antiskyrmions (aSK) emerges depending on the field strength. For all these fields,  the band structure remains metallic (to be precise topological metal), with numerous band crossings at different energies. As such, when $H$=0.4 T, an $\mathcal{RK}$ joint topological antiskyrmion ($\mathcal{RK}$-aSK) appears with a certain number of chiral boundary states (N$_{CBs}$). A $\mathcal{RK}$-aSK is characterized by the antiskyrmion surrounded by non-trivial  topologically protected chiral boundary states\cite{p67}. The positive (anticlockwise arrow, Fig.  \ref{f5}) and negative (clockwise arrow,  \ref{f5}) sign of NCBs  reflects the different chirality. When H=0.2 T, a $\mathcal{RK}$--aSK state with opposite chirality appears (see Fig.  \ref{f5}). And, in the absence of any field ($H$=0 T), the system transforms into the mixing phase with the co-existence of smaller sized aSKs and work-like pattern of spin spiral states, as a result of strong Dzyaloshinski-Moriya (DM) interaction. Similar multiple-space topological cross-over has also been observed under hydrostatic pressure on Mn$_2$Pd$_{0.5}$Ir$_{0.5}$Sn. Monolayer MnBi$_2$S$_2$Te$_2$ is yet another example where similar kinds of topological phase transition (as a result of competition between $\mathcal{R}$- and $\mathcal{K}$-space topology) has been observed\cite{p67}. }

\begin{figure}[ht]
\centering
\includegraphics[width=0.95\linewidth]{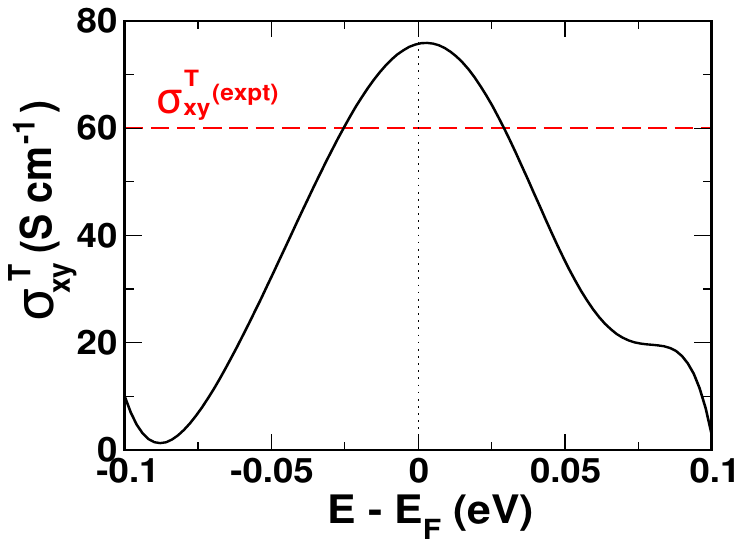}
\caption{\textcolor{black}{  The computed topological Hall conductivity, THC ($\sigma_{xy}^T$) as a function of energy (E$_F$ is the Fermi energy). The calculated THC at E$_F$, $\sigma_{xy}^T$ = 75 S cm$^{-1}$,  is broadly consistent with the measured value.}}
\label{f6}
\end{figure}

\textcolor{black}{In order to quantify the combined contribution of real space and k-space topological features, we have used the above constructed TB-Hamiltonian to compute the Berry phase, as well as the Topological Hall effect (THE) employing Kubo formula \cite{p120,p139}. The simulated Topological Hall conductivity as a function of energy for a periodic aSK (Fig. 4(b) of main manuscript) is shown in Fig. \ref{f6}. The computed topological Hall conductivity, $\sigma_{xy}^{T}$=75 S/cm (at Fermi energy E$_F$) is broadly consistent with our measured value ($\sigma_{xy}^{T}$(exp)=59.7 S/cm). We have also calculated the THE using the band structure of the lowest energy canted-1 magnetic configuration (Fig. S12(e) of SM\cite{R30}), which turns out to be 28 S/cm (at E$_F$). This clearly shows that consideration of even a simple (periodic) aSK improves the overall comparison between the simulated and measured THE. 
On a side note, it is important to mention that the measured carrier concentration (or the chemical potential) may not be the same as the Fermi energy (E$_F$) obtained from simulation, as the latter is computed for a pristine crystal. As such, a more reasonable comparison between simulated and measured $\sigma_{xy}^T$ should be made by considering a range of $\sigma_{xy}^T$ values calculated within E=E$_F \pm \delta$. } 


{\it \bf Summary}\indent To conclude, we report a combined experimental and theoretical investigation of structural, magnetic and electrical properties of novel tetragonal inverse Heusler alloy Mn$_{2}$Pd$_{0.5}$Ir$_{0.5}$Sn which reveal an exceptionally large topological Hall effect accompanied by remarkable anomalous Hall conductivity. We propose the possible unique realization of antiskyrmion in Weyl metal and discuss the potential correlation between non-trivial topological features across real and reciprocal spaces. The system shows a ferrimagnetic ordering above room temperature with spin-reorientation transition at $T = 85$ K. This results in a canted spin arrangement between the magnetic sublattices at lower temperatures that have been subsequently verified by the {\it ab-initio} calculations. The calculations confirm the presence of 42 pairs of Weyl nodes which manifest themselves in a large measured anomalous Hall conductivity of 133 S cm$^{-1}$ with an anomalous Hall angle of 3.6$\%$. Observation of a giant topological Hall effect and non-zero residual $\rho_{xy}^T$ at zero field along with significant spin-orbit coupling and crystal symmetry suggest the presence of topological magnetic spin-textures, possibly antiskyrmions. This result is corroborated with micromagnetic simulations employing experimentally obtained parameters. Our findings serve as a guiding beacon for delving deeper into the investigation of gauge field effects of Weyl fermions in Mn-based tetragonal Heusler alloys and leverage the emergent properties manifested from the relationship between reciprocal space topology and real space non-trivial magnetism for spintronic applications.

\indent
\indent
\indent
\indent
\indent
\indent

\textit{\bf{Sample Preparation and X-ray characterization}}  Polycrystalline samples were prepared by arc melting the constituent pure elements (Mn, Pd, Ir and Sn of 5N purity) in their stoichiometric ratios. The as-cast ingots were annealed at 1073 K for three days and subsequently quenched in ice water. The annealed samples were thoroughly crushed into powders and room temperature X-ray diffraction (XRD) were performed on several samples using Rigaku TTRX-III system.

\textit{\bf{TEM characterization}} TEM specimen is prepared using the standard method of mechanical thinning, dimpling and  final thinning using a precision-ion-polishing system (PIPS, Gatan, Pleasanton, CA). The ion polish was carried out at 3.0 keV energy and followed by a 1.2 keV cleaning process. TEM investigation was carried out using FEI, Tecnai G2 F30, S-Twin microscope operating at 300 kV. The compositional analysis was performed by energy dispersive X-ray spectroscopy (EDS, EDAX Instruments) attachment on the Tecnai G2 F30.

\textit{\bf{Magnetic measurements}} Magnetic measurements were performed using MPMS-3 SQUID-VSM equipped with 7 T superconducting magnets. For each isothermal magnetic field-dependent magnetization measurement (\textit{M versus H}), the system was zero field cooled from paramagnetic region to target temperature (\textit{T}) and allowed to rest for three minutes. For all magnetization measurements, the applied magnetic field was applied perpendicular to \textit{length} $\times$ \textit{breadth} area of the sample.

\textit{\bf{Magnetotransport measurements}} For magnetotransport measurements, rectangular pieces of approximate dimensions 3.5 mm $\times$ 1.5 mm $\times$ 0.42 mm were used and electrical contacts were made by $25\mu m$ gold wires using silver epoxy. For field dependent measurements, the applied magnetic field was ramped in the sequence +7T $\rightarrow$ -7T $\rightarrow$ +7T $\rightarrow$ -7T, both for magnetization and magnetotransport measurements. To avoid the demagnetization effect both magnetic and transport measurements were done on the same piece with field perpendicular to $length\times breadth$ surface area with a maximum error field of 20 Oe.

\textit{\bf{ Computational Details:}} We used generalized gradient approximation by Perdew-Burke-Ernzerhof (PBE) to capture the exchange and correlation effect. Plane-wave basis set within the projector augmented wave (PAW) method was used with an energy cutoff of 350 eV. 
A 32 atom unit cell is used to simulate the system, Mn$_{2}$Pd$_{0.5}$Ir$_{0.5}$Sn.  In order to model the (Pd,Ir) mixing, we had generated several possible ordered configurations (70 of them) and simulated their energetics. As expected, several of these configurations turn out to be degenerate. Out of the non-degenerate configurations, we chose only those that acquire the space group $I\bar{4}2d$ (SG: 122), which is the experimentally predicted structure (see Sec. II). In total, there are 8 such structural configurations, whose energetics are shown in Table ST2 of SM\cite{R30}. We chose the lowest energy configuration for doing further calculations. We had also simulated the experimentally measured structure of Mn$_2$Pd$_{0.5}$Ir$_{0.5}$Sn (see Table ST1 of SM\cite{R30}), whose energy turn out to be very close (sub-meV) to the lowest energy configuration (Configuration 1).   
In order to further check the possibility of completely random disorder between Pd and Ir, we have generated a special quasi random structure (SQS)\cite{p112} involving 32 atoms. SQS is an ordered supercell which is constructed in such a way
as to mimic the most relevant pair and multisite correlation
functions of the disordered phase.
The SQS structure is found to be 90 meV higher in energy as compared to the lowest energy structure.
Brillouin zone (BZ) integrations were performed using a 4$\times$4$\times$2 k-mesh for all the 32-atom unit cells, with total energy (force) converged up to 10$^{-6}$ eV (0.01 eV/$\AA$). The spin-orbit coupling (SOC) effect was included in all the calculations.
We have also checked the thermal and mechanical stability of Mn$_2$Pd$_{0.5}$Ir$_{0.5}$Sn by calculating the phonon dispersion and elastic constants. These results are shown in Fig. S10 and Table ST3 of  SM\cite{R30}.
Maximally localized Wannier functions (MLWF) were used to generate the low-energy tight-binding Hamiltonian. Wannier-Tools was used to simulate the topological properties e.g. chirality, Berry curvature and anomalous Hall conductivity.

\textit{Acknowledgement}\indent Bhattacharya and Ahmed would like to acknowledge SINP, India and the Department of Atomic Energy (DAE), Government of India for research funding and Fellowship. We thank Prof. Sanjeev Kumar, IISER Mohali, for the discussion. O. A. T. acknowledges support by the Australian Research Council (Grant Nos. DP200101027 and DP240101062) and by the NCMAS grant.

\bibliographystyle{unsrt}

\bibliography{ref}

\end{document}